# Partner support during pregnancy mediates social inequalities in maternal postpartum depression for non-migrant and first generation migrant women

Aurélie Nakamura[1,2,*], Fabienne El-Khoury Lesueur[1], Anne-Laure Sutter-Dallay[3,4], Jeanna-Eve Franck[1], Xavier Thierry[5], Maria Melchior[1,6], Judith van der Waerden[1]

1 Sorbonne Université, INSERM, Institut Pierre Louis d'Épidémiologie et de Santé Publique (IPLESP), Department of Social Epidemiology, F75012 Paris, France

2 French School of Public Health (EHESP), Doctoral Network, Rennes, France

3 INSERM, UMR 1219, Bordeaux Population Health, Bordeaux University, France

4 University Department of Adult Psychiatry, Charles-Perrens Hospital, 33000, Bordeaux, France.

5 UMS Elfe Team, Institut National d'Etudes Démographiques (INED), F75000 Paris, France

6 French collaborative Institute on Migration (ICM), Paris, France

**\*Corresponding author:**

Aurélie Nakamura

Sorbonne Université, INSERM, Institut Pierre Louis d'Épidémiologie et de Santé Publique (IPLESP), Department of Social Epidemiology,

Faculté de Médecine St Antoine, 27 rue de Chaligny, 75012 Paris, France

aurelie.nakamura@iplesp.upmc.fr

+33681817496

**Abstract**

**Background** An advantaged socioeconomic position (SEP) and satisfying social support during pregnancy (SSP) have been found to be protective factors of maternal postpartum depression (PDD). An advantaged SEP is also associated with satisfying SSP, making SSP a potential mediator of social inequalities in PPD. SEP, SSP and PPD are associated with migrant status. The aim of this study was to quantify the mediating role of SSP in social inequalities in PPD regarding mother's migrant status.

**Methods** A sub-sample of 15,000 mothers from the French nationally-representative ELFE cohort study was used for the present analyses. SEP was constructed as a latent variable measured with educational attainment, occupational grade, employment, financial difficulties and household income. SSP was characterized as perceived support from partner (good relation, satisfying support and paternal leave) and actual support from midwives (psychosocial risk factors assessment and antenatal education). Mediation analyses with multiple mediators, stratified by migrant status were conducted.

**Results** Study population included 76% of non-migrant women, 12% of second and 12% of first generation migrant. SEP was positively associated with support from partner, regardless of migrant status. Satisfying partner support was associated with a 8 (non-migrant women) to 11% (first generation migrant women) reduction in PPD score.

**Limitations** History of depression was not reported.

**Conclusions** Partner support could reduce social inequalities in PPD. This work supports the need of interventions, longitudinal and qualitative studies including fathers and adapted to women at risk of PPD to better understand the role of SSP in social inequalities in PPD.

**Keywords** social support, postpartum depression, epidemiology, social inequalities, pregnancy, mediation analysis

**Introduction**

Maternal postpartum depression (PPD) is defined as minor or major depression occurring within a year after giving birth (Gaynes et al., 2005). PPD impacts the mother's quality of life as well as family functioning and can have consequences for offspring development and mental health (Shorey et al., 2018). It is the most common postnatal complication, affecting between 5 and 25% of mothers in Western countries (Bérard et al., 2019; O'Hara and McCabe, 2013; Shorey et al., 2018), with prevalence rates for socioeconomically disadvantaged women estimated to be even higher (38%) (Seguin et al., 1999). Likewise, the prevalence of PPD in migrant women living in Western countries, including France, is also higher than in the general population (estimated to vary from 24% to 42%) (Collins et al., 2011). Research shows that the causes of PPD are multifactorial (Abdollahi et al., 2016; Halbreich, 2005), but the underlying causal mechanisms which precipitate the onset of depression are still poorly understood (Yim et al., 2015).

One important risk factor associated with PPD is a disadvantaged socioeconomic position (SEP) (Beck, 2001; Goyal et al., 2010; O'Hara and McCabe, 2013; Robertson et al., 2004). Associations between different aspects of SEP and depression have been studied extensively (Lancaster et al., 2010). Low income and low occupational grade may be linked to depression through income-related stressors, such as financial strain (El-Khoury et al., 2018). Low educational attainment could be linked to low health literacy and the inability to recognize symptoms of depression and seek help.

A disadvantaged SEP also tends to be positively associated with lack of perceived social support (Mangrio et al., 2011), independently from actual support (Milgrom et al., 2019). Perceived social support, defined as satisfaction with both informal (i.e. partner, family, friends and significant others) and formal (i.e. midwife, general practitioner and other health care professionals) social relations, has consistently been associated with positive health outcomes. It can be distinguished from actual social support, defined as the sum of supportive behaviors a person benefits from

(Melrose et al., 2015). Lack of social support during pregnancy (SSP) (Leahy-Warren et al., 2012; Nielsen et al., 2000; O'Hara and McCabe, 2013), and especially actual and perceived partner support during pregnancy (Milgrom et al., 2019, 2008; O'Hara, 1986; Stapleton et al., 2012), has been found to be associated with higher risk of PPD. Thus, it is possible that social inequalities in PPD can partly be attributed to SSP, as described in the theoretical framework in **Figure 1,** inspired by Milgrom et al. (Milgrom et al., 2019).

The effect of SSP on social inequalities in PPD could be even more pronounced among migrant women. When first generation migrant women leave their country of origin, they tend to be in better health than the general population of their birth country ("healthy migrant effect") (McDonald and Kennedy, 2004). However, negative events can impact migrant women's health, either during the journey or upon arrival in the host country. Additionally, migrant women tend to have a diminished SEP in their host country due to employment difficulties, administrative insecurity, or housing difficulties (Beauchemin et al., 2016; Tortelli et al., 2017), which are risk factors of poor mental health. Also, disadvantaged SEP is associated with barriers to accessing health care systems, due to language barriers, low literacy or an inability for health professionals to understand women's needs (International Organization for Migration, 2011; Schmied et al., 2017). Particular to migrant women as well might be their reduced access to social support structures, increasing their difficulties to cope with health issues (Essén et al., 2018). A lack of social support (exacerbated by loss of support from family, friends and community, difficulty of creating new social support in the host country and discrimination), disadvantaged SEP and life stressors have been noted as key mechanisms of PPD in migrant women (Saad, 2019).

To our knowledge, few studies examined the mediating role of social support in social inequalities with regard to depression in pregnant (Rahman et al., 2014; Wei et al., 2018) and postpartum migrant women (Gjerdingen et al., 2014; Rahman et al., 2014). The study conducted by Gjerdingen et al. (2014) found that postpartum employment and perceived social support were independently associated with decreased risk of PPD and in Rahman et al. (2014), low SSP mediated the association between maternal educational level and PPD. However, these studies presented both methodological and statistical limitations. In the first study, social support was assed using a composite measure that included perceived and actual social support from the partner and others but did not include formal dimensions of SSP. Furthermore, SEP was only evaluated though postpartum employment, while PPD was assessed using only a two-item questionnaire (PHQ-2) (Gjerdingen et al., 2014). In the second study, education and income were used as indicators of SEP but no distinction was made between different aspects of social support. (Rahman et al., 2014); Moreover, as social support was measured at the same time as PPD, causal inference about the relationship between these two dimensions could not be drawn. Finally, multiple mediation and moderation pathways were tested, but correlation between the different mediators was not taken into account. (Rahman et al., 2014).

To our knowledge, no studies in this area distinguished women who are first or second generation migrants. We hypothesize that first generation migrant women have lower socioeconomic circumstances as well as levels of access to health care systems and social support networks compared to second-generation and non-migrant women, due to a potentially reduced integration and knowledge of available health resources. Indeed, El-Khoury et al. (2018) showed that first generation migrant women have an increased probability of PPD but not second generation migrant women. Therefore, the aim of this study was to quantify the mediating effect of perceived and actual SSP in social inequalities in terms of maternal PPD according to women's migrant status.

**Methods**

*The ELFE cohort study*

The ELFE study (*Etude Longitudinale Française depuis l'Enfance*) is a birth cohort that recruited 18,329 children representative of children born in France in 2011 in 320 maternity wards using random sampling (Charles et al., 2019). ELFE aims to follow children from birth to adulthood.

Inclusion criteria were: singletons or twins born after at least 33 weeks gestation, mothers of at least 18 years of age and not planning on moving outside of Metropolitan France in the three years following study inclusion. Mothers had to be able to give consent either in French, English, Arabic or Turkish. Data were collected at birth via face-to-face interviews conducted by midwives and by self-reported questionnaires. Information on maternal depression were collected by phone interviews when the child was two months, one year and two years old. The ELFE study received the approval of France's bodies regulating ethical research conduct (Comité Consultatif sur le Traitement des Informations pour la Recherche en Santé: CCTIRS; Commission National Informatique et Libertés: CNIL). The population for the current study consisted of 14,587 women who gave birth to singletons and who had complete data on postnatal depression scores, SEP, SSP, migrant status and other covariates. (Figure 2).

**Measures**

*Postpartum depression (PPD)*

Postpartum depressive symptoms were assessed at two months after child birth using the Edinburgh Postnatal Depression Scale (EPDS) (Cox et al., 1987). The EPDS scale comprises ten items concerning the past week; each question having four possible answers, ranked from 0 to 3 points. Thus, the total EPDS score ranges from 0 to 30 points. The EPDS scale was validated for French women (Adouard et al., 2005; Guedeney and Fermanian, 1998). Cronbach's alpha for our study sample was 0.80. In the analyses, EPDS score was used as a continuous variable.

*Socioeconomic position (SEP)*

Socioeconomic position was measured through five indicators: 1) maternal educational attainment (< high school diploma; high school diploma; two-year post high school degree; >two-year post high school degree), 2) maternal occupational grade (none; low (e.g. clerk, manual worker); intermediate (e.g. mid-level manager, technician); high (e.g. manager)), 3) maternal employment during pregnancy (no; yes), 4) household income weighted by the number of people residing in the household (Labrador, 2013) and 5) household financial difficulties during pregnancy (yes; no).

*Social support during pregnancy (SSP)*

SSP was assessed using informal dimensions of perceived social support and formal dimensions of actual social support. Informal support was characterized by 1) good relations with the partner (frequent quarrels with the partner: no; yes), 2) satisfying partner support (no; yes) and 3) paternity leave (already taken or intents to take it: no; yes). Formal support was assessed by 1) early prenatal psychosocial risk factors assessment (no; yes) and 2) number of antenatal classes. Characterization of informal and formal social support in this study was described in another article (Nakamura et al., 2020). Briefly, early prenatal psychosocial risk factors assessment is supposed to be offered to all expected parents by a midwife. It is a 45 minutes appointment during which future parents have the opportunity of addressing their concerns regarding pregnancy's social environment, psychological difficulties and other concerns (Isserlis et al., 2008).

*Migrant status*

Classification of mother's migrant status was detailed in a previous article. (El-Khoury et al., 2018). Briefly, women's migrant status was categorized as 1) non-migrant women (French born to French parents), 2) second generation migrant (French with at least one immigrant parent) and 3) first generation migrant (immigrant).

C*ovariates*

Potential confounders of the association between SEP and PPD, SEP and SSP or SSP and PPD included: mother's and father's age at birth (continuous), parity (0; ≥1 other child), marital status (married; in civil union; other), partner's employment status (yes; no), timing of pregnancy (satisfying: yes; no), prior postpartum depression (yes; no), psychological difficulties during pregnancy (yes; no), offspring's sex (girl; boy) and breastfeeding at birth (exclusive; non-exclusive; no).

Statistical analyses

This study aims to quantify the mediating effect of perceived SSP in social inequalities of maternal PPD with regard to women's migrant status, First, SEP was built as a latent variable using confirmatory factorial analysis (CFA), explaining highest level of education attained, occupational grade, employment during pregnancy, household income and financial difficulties during pregnancy (**Supplementary Figure 1**). Associations between 1) SEP and aspects of SSP, 2) SEP and PPD symptoms and 3) aspects of SSP and PPD were assessed using bivariate and multivariate linear (PPD) or logistic (SSP) regressions depending on the outcome. Then, these analyses were combined in a mediation analysis in order to estimate i) the natural direct effect of SEP on PPD ii) the natural indirect effects and proportions mediated of SEP on PPD, passing through the five aspects of SSP identified (Muthén and Asparouhov, 2015). Total effect of SEP on PPD was defined as the sum of direct and indirect effects. As some of the mediating variables were correlated, we used mediation modeling based on Nguyen et al. (Nguyen et al., 2016). Goodness of fit of mediation models was assessed using root mean square error of approximation (RMSEA), with a value of 0.06 or lower indicating an acceptable model fit and comparative fit index (CFI) and Tucker-Lewis Index (TLI), both examining the discrepancy between data and hypothesized model, with a value of 0.90 or greater indicating an acceptable fit and a value greater or equal to 0.95 indicating a good fit (Hu and Bentler, 1999).

Despite the inclusion of previously indicated covariates, we cannot exclude that some potential confounders were unmeasured or not controlled for. In order to test the robustness of our analyses, we conducted sensitivity analyses by calculating e-values i.e. the strengths of association between at least one unmeasured confounders and SEP, SSP and PPD that would make tested associations statistically non-significant (Ding and VanderWeele, 2016; VanderWeele and Ding,

2017). E-values are presented in odds ratio (OR) or relative risk (RR) scale. Mediation analyses were conducted using MPLUS 8.1 (Muthén and Muthén) and e-values were calculated using R package EValue (Mathur et al., 2018).

**Results**

*Characteristics of study participants*

Second generation migrants represented twelve percent of women in our sample (95% from European Union). The same proportion of women were first generation migrants, including 15% of women from European Union, 40% from North Africa, 23% of women from Sub-Saharan Africa and 22% from other regions. The average EPDS score was 5.56 (+/- 4.39) for non-migrant women, 6.03 (+/- 4.60) for second generation migrant women and 7.70 (+/- 4.77) for first generation migrant women (**Table 1**).

SEP was associated with migrant status: non-migrant women had a better SEP than second generation migrant women, which in turn had a better SEP than first generation migrant women Among non-migrant women, 38% had a diploma higher than a two-year degree (vs. 31% for second generation migrant women and 30% for first generation migrant women) and 12% of non-migrant women had an educational attainment lower than high school (vs. 17% for women from the second migrant generation and 21% for women from the first migrant generation). More than four out of five women from the non-migrant and second migrant generation worked during their pregnancy vs. one out of two women from the first migrant generation and half of the women had a low occupational grade, regardless of their migrant status. Forty three percent of non-migrant women declared having had financial difficulties during their pregnancy (vs. 47% of women from the second migrant generation and 52% of women from the first migrant generation). The average household income per unit of consumption of non-migrant women was 1,674 euros/month (vs. 1,581 euros/month for women from the second migrant generation and 1,395 euros/month for women from the first migrant generation). Women were on average 31 years old at childbirth, half of them were multiparous and more than 90% of women were in a partnership, regardless their migrant status (**Table 1**).

Most women felt satisfied with the social support they received from their partner during pregnancy (90% for non-migrant women vs. 86% and 84% for women from the second and first migrant generation respectively) and had a good relationship with their partner during pregnancy. More than two thirds of partners took a paternity leave (79% for non-migrant women vs. 71% and 65% for second and first generation migrant women). However, formal support through psychosocial risk assessment and antenatal education was particularly low in first generation migrant women in comparison to non-migrant women (26% vs. 36% of women attended a prenatal psychosocial risk factor assessment and on average 2 vs. 4 antenatal classes).

*Mediation analyses between socioeconomic position, social support during pregnancy and postpartum depression, stratified by migrant status*

Informal and formal social support variables were first tested in five single mediator models not stratified on migrant status. Good relationships with the partner, satisfying partner support and antenatal education mediated the relation between SEP and EPDS score, and only these three variables were included in multiple mediators model. SEP was negatively associated with EPDS score, regardless of mother's migrant status, with an increase of one unit of SEP associated with a reduction of respectively 6%, 10% and 16% of EPDS score in non-migrant women (RR = 0.94 [95%CI 0.91-0.96]), second generation migrant women (RR = 0.90 [95%CI 0.86-0.96]) and first generation migrant women (RR = 0.84 [95%CI 0.76-0.95]). SEP was also positively associated with good partner relationships in non-migrant women (OR = 1.05 [95%CI 0.99-1.12]) and, regardless of women's migrant status, with sufficient partner support (especially in women from second migrant generation: OR = 1.35 [95%Ci 1.21-1.51]) and antenatal education (especially with women from the first migrant generation: RR = 1.50 [95%CI 1.41-1.60]). Good partner relationship was associated with reduced EPDS scores of respectively 17% in non-migrant women (RR = 0.83 [95% CI 0.79-0.87]), 16% in second generation migrant women (RR = 0.84 [95% CI 0.72-0.97]) and 11% in first generation migrant women (RR = 0.89 [95% CI 0.78-1.01]). Satisfactory partner support was also associated with lower EPDS scores of respectively 7% in non-migrant women (RR = 0.93 [95%CI 0.89-0.97]) and 11% in first generation migrant women (RR = 0.89 [95%CI 0.81-0.98]) but not in second generation migrant women. An additional antenatal class attendance was associated with an increase of 5% of EPDS score in non-migrant women (RR = 1.05 [95%CI 1.03-1.08]), but not in second and first generation migrant women (**Figure 3**).

The effects of an increase of SEP on EPDS score were mediated by good partner relationship (6%), satisfying partner support (8%) and the number of antenatal classes (-9%) in non-migrant women, leading to a small total natural indirect effect of social support during pregnancy in the association between SEP and EPDS score (proportion of social inequalities of PPD symptoms mediated by the three aspects of SSP = 5%) (**Table 2**). In women from the first generation migration, the effects of an increase of SEP on EPDS score were related to satisfying partner support (11%) (**Table 2**). As there was no association at the same time between i) SEP and SSP and ii) SSP and PPD, no proportion mediated was estimated in second generation migrant women.

*Sensitivity analyses to unmeasured confounders*

Sensitivity analyses to evaluate unmeasured or uncontrolled confounding factors in non-migrant women resulted in E-Values close to 1 for the associations between SEP and good partner relationship and antenatal education and EPDS score, weakening the hypothesis of a causal relationship between these variables (respectively E-Value = 1.18 [95%CI 1.00, +inf[ and E-Value = 1.28 [95%CI 1.21, +inf[). However, as the E-Values for the associations between SEP, satisfying partner support and EPDS score were relatively high in non-migrant women in relation to the estimated RRs in this study, it is unlikely that the impact of uncontrolled for or unmeasured confounding factors would have unduly changed the main results from these analyses (**Supplementary Table 1**).

**Discussion**

*Main results*

We aimed to quantify the mediating effect of perceived SSP in social inequalities of maternal PPD using data from a nationally representative cohort of mothers giving birth in France. To our knowledge, this is the first study that used several aspects of SSP as a potential mechanism of social inequalities in PPD according to women's migrant status. A higher socioeconomic position was directly associated with lower postpartum depression risk, but also mediated through good partner relationships and satisfying partner support during pregnancy. When looking specifically at migrant women's social support, only sufficient partner support during pregnancy mediated social inequalities in PPD by 11% for first generation migrant women.

*Plausible pathways between socioeconomic position (SEP), social support during pregnancy (SSP) and postpartum depression (PPD)*

In accordance with the extensive literature on psychosocial risk factors of PPD, we found that higher SEP and perceived social support from the partner during pregnancy are associated with a reduction in the risk of PPD (Gjerdingen et al., 2014; O'Hara and McCabe, 2013; Rahman et al., 2014). On the other hand, disadvantaged SEP and lack of SSP are both well known risk factors for distress during pregnancy, which is associated with an increased risk of PPD (O'Hara and McCabe, 2013). Socioeconomically disadvantaged women had lower levels of informal support from their partner. Women with low SEP are more likely to be single parents or to have lower informal social support from their partner (Goyal et al., 2010). Thus, for married women partner support and relationships within the couple might be one of the main components of social support protective factors of PPD. However, single women might have other sources of social support during pregnancy, such as friends and family support, which were not measured in our study (Reid and Taylor, 2015).

However, contrary to previous studies, we did not find that formal social support was associated with reduced postpartum depression risk (Leahy-Warren et al., 2012). Surprisingly, the number of attended antenatal education classes was associated with an increase in the PPD symptoms, leading to a negative mediated proportion of social inequalities of PPD linked to antenatal education. Previous studies using the ELFE cohort showed that less than 40% of women received a 45 minutes of psychosocial risk factors assessment, and attended on average only three antenatal education classes (Barandon et al., 2016) . Moreover, practices regarding psychosocial factors assessment are not codified, which could introduce heterogeneity on its benefits against PPD (Barandon et al., 2016). One other plausible explanation is that especially women diagnosed with depression prior to or during pregnancy do not systematically attend antenatal classes. As mentioned by Leahy-Warren et al., (2012) limited time spent with health professionals during pregnancy could explain the lack of a significant protective effect of formal support in PPD.

Women with a lower SEP have been reported to attend less antenatal education than other women (Milcent and Zbiri, 2018) and perinatal health care (Linard et al., 2018). Among barriers to access health care, socioeconomically disadvantaged pregnant women cite absence of child care, fatigue, long waiting times and overcrowding in the clinic (Hansotte et al., 2017; Loveland Cook et al., 1999; Padilla et al., 2016). This particularly applies to first generation migrant women, who are more likely to experience socioeconomic disadvantage as well as migration-related stressors. Even if they often have a better health than non-migrant women before migration ("healthy migrant effect"), migrant women face difficulties to find an accommodation and employment, (Tortelli et al., 2017) in addition of being more at risk of having difficulties with speaking the language of the host country.

When they migrate, women also lose social support. In a meta-ethnographic study including 12 studies and 256 migrant women, women frequently reported feeling alone and worried about themselves and their children, loss of family, friends and community social support in the host country (Schmied et al., 2017). Migrant women also reported a fear of judgment and feeling of being a bad mother, disappointing the partner and shame (Schmied et al., 2017). Moreover, women perceived their emotional distress as originating from their disadvantaged SEP and not PPD (Hansotte et al., 2017; Schmied et al., 2017). In a Swedish study including 3,000 women among which 10% of women were non-Swedish speakers (which can be view as a proxy for migrant women), physical care (related to breastfeeding or coping with labor) was considered by the women more important than psychological care (Fabian et al., 2008). For some women, PPD was not recognized in their culture and they felt reluctant talking to health professionals because of humiliation and stigma associated with mental illness (Schmied et al., 2017).

*Strengths and limitations*

First, with a large sample of +14,000 women in complete case analyses, the ELFE cohort study benefited of a large sample size compared to previous studies on this topic, permitting stratified analyses by women's migrant status. Also, repeated data collection across the perinatal period, allowed us to respect temporality between SEP, SSP and PPD measurements and to include several potential confounding factors in our mediation analyses.

Second, by using latent variables analyses we were able to include different dimensions of SEP. Third, using rigorous multiple mediators modeling, we could assess the contribution of different dimensions of social support in social inequalities of PPD.

However, some limitations need to be addressed. First, important confounding factors (history of depression, experience of stressful events prior or during pregnancy (O'Hara and McCabe, 2013)) were not measured. Second, assessment of informal support during pregnancy was limited by the absence of a validated scale to assess informal and formal dimensions of SSP. In particular, support from other members of the family, friends, and women's network were not measured (Mlotshwa et al., 2017). Moreover, only qualitative aspects of women's social support (e.g. quality of the relationship with the partner) vs. quantitative aspects (e.g. number of close friends) were assessed. Third, more than 13% of women were lost to follow up at two months after birth. These women are more likely to have a lower SEP and less SSP (Nakamura et al., 2020). Finally, our mediation analyses, especially according to migrant status, tented to lack statistical power, which may have led to underestimating associations between SEP and SSP and SSP and PPD, resulting in non-significant associations between these variables.

*Recommendations for practice*

Many studies have shown the impact of depression during the perinatal period on mothers, family relationships but also on children's health (Ahun et al., 2018; Gutierrez-Galve et al., 2019). A recent systematic review for the US Preventive Services Task force (O'Connor et al., 2019) recommended counseling interventions (including cognitive behavioral therapy, interpersonal psychotherapy but also peer-based therapy) for preventing perinatal depression. Home-visit interventions, administered by trained nurses, are especially interesting for low SEP women that have more physical barriers to access health care and who can be more isolated have also shown some benefits for preventing PPD (Hansotte et al., 2017).

In a systematic review, Pilkington et al. (Pilkington et al., 2015) reported some evidence for benefits of partner-inclusive interventions for preventing PPD by strengthening couple relationship (Shapiro and Gottman, 2005), knowledge about birth and life adjustments to being new parents (Matthey et al., 2004).

Interventions targeting socioeconomically disadvantaged women, who are at higher risk of PPD (Pilkington et al., 2015), conducted by teams of nurses or midwives, social workers and peers should be developed more. Indeed, women living in precarious conditions are more concerned by finding a house or food than their mental health condition (Beeber et al., 2004). A team work could thus provide a better adhesion in the intervention from the women.

**Conclusion**

Partner support during pregnancy, especially a good relationship with the partner, is linked to reduced social inequalities of postpartum depression, both in non-migrant and in first-generation migrant women. To clarify the role of social support during pregnancy in social inequalities in postpartum depression, especially for migrant women, further studies including more disadvantaged populations would be needed. In addition, qualitative studies could provide more insights in how women define and perceive social support during pregnancy, regarding their social and socioeconomic position in order to be able to develop better interventions. More studies before and during pregnancy involving partners and adapted to characteristics of women at risk (such as living conditions, language spoken and cultural and psychiatric background) are needed to evaluate the role of informal and formal social for preventing postpartum depression.

**Acknowledgements**

The Elfe survey is a joint project between the French Institute for Demographic Studies (INED) and the National Institute of Health and Medical Research (INSERM), in partnership with the French blood transfusion service (Etablissement français du sang, EFS), Santé publique France, the National Institute for Statistics and Economic Studies (INSEE), the Direction générale de la santé (DGS, part of the Ministry of Health and Social Affairs), the Direction générale de la prévention des risques (DGPR, Ministry for the Environment), the Direction de la recherche, des études, de l'évaluation et des statistiques (DREES, Ministry of Health and Social Affairs), the Département des études, de la prospective et des statistiques (DEPS, Ministry of Culture), and the Caisse nationale des allocations familiales (CNAF), with the support of the Ministry of Higher Education and Research and the Institut national de la jeunesse et de l'éducation populaire (INJEP). Via the RECONAI platform, it receives a government grant managed by the National Research Agency under the "Investissements d'avenir" programme (ANR-11-EQPX-0038)

**References**

Abdollahi, F., Lye, M.-S., Zarghami, M., 2016. Perspective of postpartum depression theories: A narrative literature review. North American journal of medical sciences 8, 232.

Adouard, F., Glangeaud-Freudenthal, N.M.C., Golse, B., 2005. Validation of the Edinburgh postnatal depression scale (EPDS) in a sample of women with high-risk pregnancies in France. Archives of Women's Mental Health 8, 89–95. https://doi.org/10.1007/s00737-005-0077-9

Ahun, M.N., Consoli, A., Pingault, J.-B., Falissard, B., Battaglia, M., Boivin, M., Tremblay, R.E., Côté, S.M., 2018. Maternal depression symptoms and internalising problems in the offspring: the role of maternal and family factors. European child & adolescent psychiatry 27, 921–932.

Barandon, S., Bales, M., Melchior, M., Glangeaud-Freudenthal, N., Pambrun, E., Bois, C., Verdoux, H., Sutter-Dallay, A.-L., 2016. Entretien prénatal précoce et séances de préparation à la naissance et à la parentalité: caractéristiques psychosociales et obstétricales associées chez les femmes de la cohorte ELFE. Journal de Gynécologie Obstétrique et Biologie de la Reproduction 45, 599–607.

Beauchemin, C., Hamel, C., Simon, P., 2016. Trajectoires et origines: enquête sur la diversité des populations en France. Ined éditions.

Beck, C.T., 2001. Predictors of Postpartum Depression: An Update. Nursing Research 50.

Beeber, L., Holditch-Davis, D., Belyea, M., Funk, S., Canuso, R., 2004. In-home intervention for depressive symptoms with low-income mothers of infants and toddlers in the United States. Health care for women international 25, 561–580.

Bérard, A., Abbas-Chorfa, F., Kassai, B., Vial, T., Nguyen, K.A., Sheehy, O., Schott, A.-M., 2019. The French Pregnancy Cohort: Medication use during pregnancy in the French population. PloS one 14, e0219095.

Charles, M.-A., Thierry, X., Lanoe, J.-L., Bois, C., Dufourg, M.N., Popa, R., Cheminat, M., Zaros, C., Geay, B., 2019. Cohort Profile: The French National cohort of children ELFE: birth to 5 years. International journal of epidemiology.

Collins, C.H., Zimmerman, C., Howard, L.M., 2011. Refugee, asylum seeker, immigrant women and postnatal depression: rates and risk factors. Archives of women's mental health 14, 3–11.

Cox, J.L., Holden, J.M., Sagovsky, R., 1987. Detection of postnatal depression. Development of the 10-item Edinburgh Postnatal Depression Scale. The British journal of psychiatry 150, 782–786.

Ding, P., VanderWeele, T.J., 2016. Sensitivity analysis without assumptions. Epidemiology (Cambridge, Mass.) 27, 368.

El-Khoury, F., Sutter-Dallay, A.-L., Panico, L., Charles, M.-A., Azria, E., Van der Waerden, J., Melchior, M., 2018. Women's mental health in the perinatal period according to migrant status: the French representative ELFE birth cohort. The European Journal of Public Health 28, 458–463.

Essén, B., Puthooparambil, S.J., Mosselmans, L., Salzmann, T., 2018. Improving the health care of pregnant refugee and migrant women and newborn children: Technical guidance.

Fabian, H., Rådestad, I., Rodriguez, A., Waldenström, U., 2008. Women with non-Swedish speaking background and their children: a longitudinal study of uptake of care and maternal and child health. Acta paediatrica 97, 1721–1728.

Gaynes, B.N., Gavin, N., Meltzer-Brody, S., Lohr, K.N., Swinson, T., Gartlehner, G., Brody, S., Miller, W.C., 2005. Perinatal depression: Prevalence, screening accuracy, and screening outcomes: Summary.

Gjerdingen, D., McGovern, P., Attanasio, L., Johnson, P.J., Kozhimannil, K.B., 2014. Maternal depressive symptoms, employment, and social support. J Am Board Fam Med 27, 87–96.

Goyal, D., Gay, C., Lee, K.A., 2010. How Much Does Low Socioeconomic Status Increase the Risk of Prenatal and Postpartum Depressive Symptoms in First-Time Mothers? Women's Health Issues 20, 96–104

Guedeney, N., Fermanian, J., 1998. Validation study of the French version of the Edinburgh Postnatal Depression Scale (EPDS): new results about use and psychometric properties. European Psychiatry 13, 83–89.

Gutierrez-Galve, L., Stein, A., Hanington, L., Heron, J., Lewis, G., O'Farrelly, C., Ramchandani, P.G., 2019. Association of maternal and paternal depression in the postnatal period with offspring depression at age 18 years. JAMA psychiatry 76, 290–296.

Halbreich, U., 2005. Postpartum disorders: multiple interacting underlying mechanisms and risk factors. Journal of Affective Disorders 88, 1–7.

Hansotte, E., Payne, S.I., Babich, S.M., 2017. Positive postpartum depression screening practices and subsequent mental health treatment for low-income women in Western countries: a systematic literature review. Public Health Reviews 38, 3.

Hu, L., Bentler, P.M., 1999. Cutoff criteria for fit indexes in covariance structure analysis: Conventional criteria versus new alternatives. Structural equation modeling: a multidisciplinary journal 6, 1–55.

International Organization for Migration, 2011. Glossary on Migration. International Migration Law Series 25.

Isserlis, C., Sutter-Dallay, A., Glangeaud-Freudenthal, N., 2008. Guide pour la pratique de l'entretien prénatal précoce. Toulouse, érès.

Labrador, J., 2013. Une forte hétérogénéité des revenus en Ile-de-France.

Lancaster, C.A., Gold, K.J., Flynn, H.A., Yoo, H., Marcus, S.M., Davis, M.M., 2010. Risk factors for depressive symptoms during pregnancy: a systematic review. American Journal of Obstetrics and Gynecology 202, 5–14. https://doi.org/10.1016/j.ajog.2009.09.007

Leahy-Warren, P., McCarthy, G., Corcoran, P., 2012. First-time mothers: social support, maternal parental self-efficacy and postnatal depression. Journal of clinical nursing 21, 388–397.

Linard, M., Blondel, B., Estellat, C., Deneux-Tharaux, C., Luton, D., Oury, J., Schmitz, T., Mandelbrot, L., Azria, E., PreCARE Study Group, 2018. Association between inadequate antenatal care utilisation and severe perinatal and maternal morbidity: An analysis in the Pre CARE cohort. BJOG: An International Journal of Obstetrics & Gynaecology 125, 587–595.

Loveland Cook, C.A., Selig, K.L., Wedge, B.J., Gohn-Baube, E.A., 1999. Access barriers and the use of prenatal care by low-income, inner-city women. Social Work 44, 129–139.

Mangrio, E., Hansen, K., Lindström, M., Köhler, M., Rosvall, M., 2011. Maternal educational level, parental preventive behavior, risk behavior, social support and medical care consumption in 8-month-old children in Malmö, Sweden. BMC public health 11, 891.

Mathur, M.B., Ding, P., Riddell, C.A., VanderWeele, T.J., 2018. Web site and R package for computing E-values. Epidemiology 29, e45–e47.

Matthey, S., Kavanagh, D.J., Howie, P., Barnett, B., Charles, M., 2004. Prevention of postnatal distress or depression: an evaluation of an intervention at preparation for parenthood classes. Journal of Affective Disorders 79, 113–126.

McDonald, J.T., Kennedy, S., 2004. Insights into the 'healthy immigrant effect': health status and health service use of immigrants to Canada. Social science & medicine 59, 1613–1627.

Melrose, K.L., Brown, G.D., Wood, A.M., 2015. When is received social support related to perceived support and well-being? When it is needed. Personality and Individual Differences 77, 97–105.

Milcent, C., Zbiri, S., 2018. Prenatal care and socioeconomic status: effect on cesarean delivery. Health economics review 8, 7.

Milgrom, J., Gemmill, A.W., Bilszta, J.L., Hayes, B., Barnett, B., Brooks, J., Ericksen, J., Ellwood, D., Buist, A., 2008. Antenatal risk factors for postnatal depression: a large prospective study. Journal of affective disorders 108, 147–157.

Milgrom, J., Hirshler, Y., Reece, J., Holt, C., Gemmill, A.W., 2019. Social Support—A Protective Factor for Depressed Perinatal Women? International journal of environmental research and public health 16, 1426.

Mlotshwa, L., Manderson, L., Merten, S., 2017. Personal support and expressions of care for pregnant women in Soweto, South Africa. Global health action 10, 1363454.

Muthén, B., Asparouhov, T., 2015. Causal effects in mediation modeling: An introduction with applications to latent variables. Structural Equation Modeling: A Multidisciplinary Journal 22, 12–23.

Nakamura, A., Sutter-Dallay, A.-L., El-Khoury Lesueur, F., Thierry, X., Gressier, F., Melchior, M., van der Waerden, J., 2020. Informal and professional social support during pregnancy and joint parental postnatal depression. The French representative ELFE cohort study. International Journal of Social Psychiatry (in press).

Nguyen, T.Q., Webb-Vargas, Y., Koning, I.M., Stuart, E.A., 2016. Causal mediation analysis with a binary outcome and multiple continuous or ordinal mediators: Simulations and application to an alcohol intervention. Structural equation modeling: a multidisciplinary journal 23, 368–383.

Nielsen, D., Videbech, P., Hedegaard, M., Dalby, J., Secher, N.J., 2000. Postpartum depression: identification of women at risk. BJOG: An International Journal of Obstetrics & Gynaecology 107, 1210–1217.

O'Connor, E., Senger, C.A., Henninger, M.L., Coppola, E., Gaynes, B.N., 2019. Interventions to prevent perinatal depression: evidence report and systematic review for the US Preventive Services Task Force. Jama 321, 588–601.

O'Hara, M.W., 1986. Social support, life events, and depression during pregnancy and the puerperium. Archives of general psychiatry 43, 569–573.

O'Hara, M.W., McCabe, J.E., 2013. Postpartum Depression: Current Status and Future Directions. Annu. Rev. Clin. Psychol. 9, 379–407. https://doi.org/10.1146/annurev-clinpsy-050212-185612

Padilla, C.M., Kihal-Talantikit, W., Perez, S., Deguen, S., 2016. Use of geographic indicators of healthcare, environment and socioeconomic factors to characterize environmental health disparities. Environmental health 15, 79.

Pilkington, P.D., Whelan, T.A., Milne, L.C., 2015. A review of partner-inclusive interventions for preventing postnatal depression and anxiety. Clinical Psychologist 19, 63–75.

Rahman, K., Bowen, A., Muhajarine, N., 2014. Examining the factors that moderate and mediate the effects on depression during pregnancy and postpartum. J Pregnancy Child Health 1, 2.

Reid, K.M., Taylor, M.G., 2015. Social support, stress, and maternal postpartum depression: A comparison of supportive relationships. Social Science Research 54, 246–262.

Robertson, E., Grace, S., Wallington, T., Stewart, D.E., 2004. Antenatal risk factors for postpartum depression: a synthesis of recent literature. General Hospital Psychiatry 26, 289–295. https://doi.org/10.1016/j.genhosppsych.2004.02.006

Saad, M., 2019. Examining the Social Patterning of Postpartum Depression by Immigration Status in Canada: an Exploratory Review of the Literature. Journal of racial and ethnic health disparities 6, 312–318.

Schmied, V., Black, E., Naidoo, N., Dahlen, H.G., Liamputtong, P., 2017. Migrant women's experiences, meanings and ways of dealing with postnatal depression: A meta-ethnographic study. PloS one 12, e0172385.

Seguin, L., Potvin, L., St-Denis, M., Loiselle, J., 1999. Depressive symptoms in the late postpartum among low socioeconomic status women. Birth 26, 157–163.

Shapiro, A.F., Gottman, J.M., 2005. Effects on marriage of a psycho-communicative-educational intervention with couples undergoing the transition to parenthood, evaluation at 1-year post intervention. The Journal of Family Communication 5, 1–24.

Shorey, S., Ing, C.C.Y., Ng, E.D., Huak, C.Y., San, W.T.W., Seng, C.Y., 2018. Prevalence and incidence of postpartum depression among healthy mothers: A systematic review and meta-analysis. Journal of psychiatric research.

Stapleton, L.R.T., Schetter, C.D., Westling, E., Rini, C., Glynn, L.M., Hobel, C.J., Sandman, C.A., 2012. Perceived partner support in pregnancy predicts lower maternal and infant distress. Journal of Family Psychology 26, 453.

Tortelli, A., Skurnik, N., Szöke, A., Simon, P., 2017. L'importance de la recherche épidémiologique psychiatrique sur les populations migrantes en France. Presented at the Annales Médico-psychologiques, revue psychiatrique, Elsevier, pp. 577–582.

VanderWeele, T.J., Ding, P., 2017. Sensitivity analysis in observational research: introducing the E-value. Annals of internal medicine 167, 268–274.

Wei, D.-M., Yeung, S.L.A., He, J.-R., Xiao, W.-Q., Lu, J.-H., Tu, S., Chen, N.-N., Lam, K.B.H., Cheng, K.-K., Leung, G.M., 2018. The role of social support in family socio-economic disparities in depressive symptoms during early pregnancy: Evidence from a Chinese birth cohort. Journal of affective disorders 238, 418–423.

Yim, I.S., Tanner Stapleton, L.R., Guardino, C.M., Hahn-Holbrook, J., Dunkel Schetter, C., 2015. Biological and psychosocial predictors of postpartum depression: systematic review and call for integration. Annual review of clinical psychology 11, 99–137.

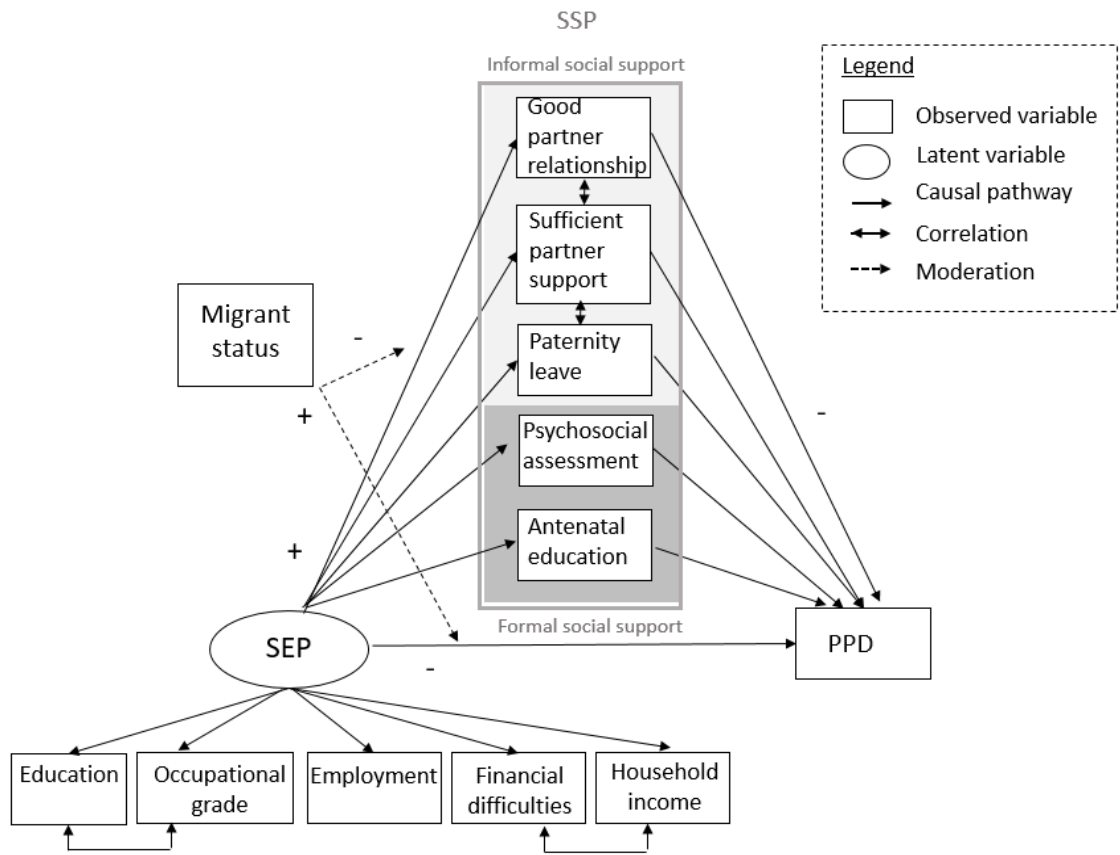

**Figure 1 Theoretical framework for the relation between socioeconomic position, informal and formal social support during pregnancy and postpartum depression**

SEP = socioeconomic position (exposure), SSP = perceived social support during pregnancy (potential mediator), PPD = postpartum depression (outcome). Potential confounders of the association between SEP and PPD, SEP and SSP or SSP and PPD included: mother's and father's age at birth, parity, marital status, partner's employment status, timing of pregnancy, prior postpartum depression, psychological difficulties during pregnancy, offspring's sex and breastfeeding at birth.

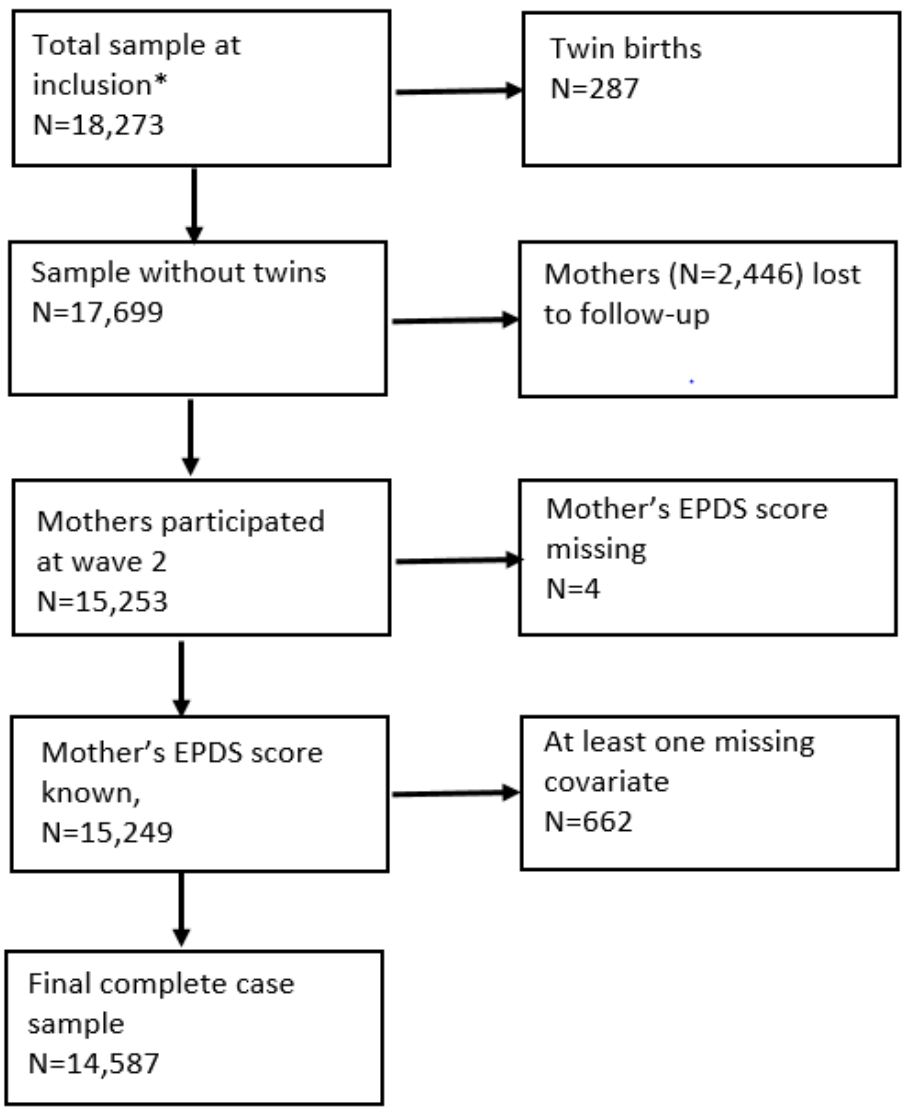

**Figure 2 – Flowchart describing sample selection, ELFE cohort study 2011-2013**

*Participants who declined their wish to stay in the study were excluded from the analyses

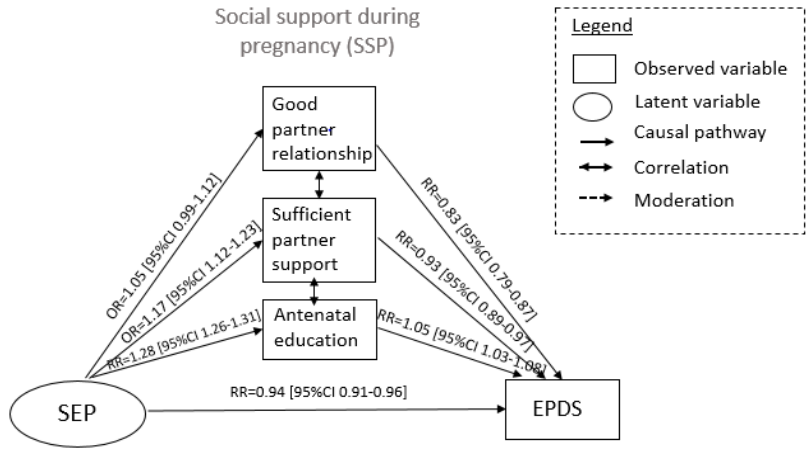

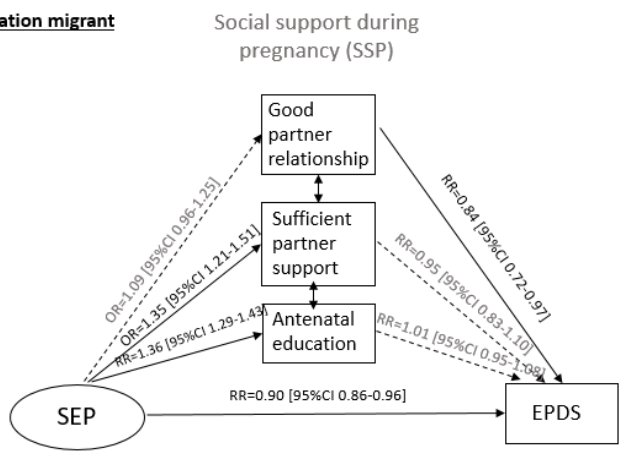

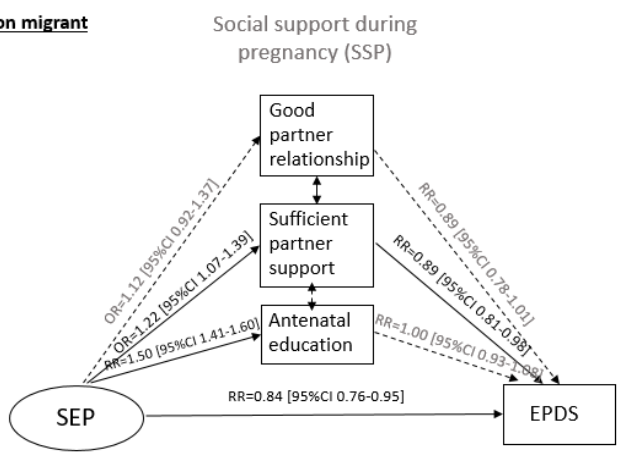

SEP = socioeconomic position, NS = not significantly different from 0, OR= Odds Ratio, RR= Relative Risk

Goodness of fit: RMSEA=0.04 [90%CI 0.04-0.04]; CFI=0.95 / TLI=0.94; WRMR=4.46;

N=11,082 (A), N=1,755 (B), N=1,750 (C)

Adjustment for mother's and father's age at child birth, parity, marital status, father's employment status, timing of pregnancy, psychological difficulties during pregnancy, previous postpartum depression, number of prenatal visits, child sex and breastfeeding at birth.

**Figure 3 Adjusted odds ratios, risk ratios, direct and indirect effect and proportion mediated of social support during pregnancy in the relation between socioeconomic status and postpartum depression, stratified by migrant status, ELFE cohort study 2011-2013**

**Table 1 – Characteristics of the sample by migrant status (N and % or means and standard deviation), ELFE cohort study 2011-2013**

|  | Non-migrant N = 11,082 (76%) | | Second generation migrant N = 1,755 (12%) | | First generation migrant N = 1,750 (12%) | | Chi2 or Welch p-value |
|---|---|---|---|---|---|---|---|
|  | N (or mean) | % (or sd) | N (or mean) | % (or sd) | N (or mean) | % (or sd) |  |
| **Postpartum depression symptoms (PPD)** | | | | | | | |
| *EPDS (continuous)* | *5.56* | *4.39* | *6.03* | *4.60* | *7.70* | *4.77* | *< 0.001* |
| **Socioeconomic position (SEP)** | | | | | | | |
| Educational attainment | | | | | | | *< 0.001* |
|     < high school | 1,592 | 14 | 303 | 17 | 368 | 21 | |
|     high school | 1,999 | 18 | 363 | 21 | 434 | 25 | |
|     Two-year degree | 3,241 | 29 | 539 | 31 | 417 | 24 | |
|     > Two-year degree | 4,250 | 38 | 550 | 31 | 531 | 30 | |
| Occupational grade | | | | | | | *< 0.001* |
|     None | 259 | 2 | 462 | 3 | 348 | 20 | |
|     Low | 5,523 | 50 | 983 | 56 | 846 | 48 | |
|     Intermediate | 3,225 | 29 | 427 | 24 | 313 | 18 | |
|     High | 2,075 | 19 | 283 | 16 | 243 | 14 | |
| Employment during pregnancy (yes) | 9,439 | 86 | 1,382 | 80 | 901 | 54 | *< 0.001* |
| Financial difficulties during pregnancy (no) | 6,298 | 57 | 908 | 53 | 642 | 48 | *< 0.001* |
| *Household income (per consumption unit in euros)* | *1,674* | *795* | *1,581* | *797* | *1,396* | *935* | *< 0.001* |
| **Perceived social support during pregnancy (SSP)** | | | | | | | |
| Satisfying partner support (yes) | 9,693 | 90 | 1,415 | 86 | 1,094 | 84 | *< 0.001* |
| Good partner relationship (yes) | 10,157 | 96 | 1,482 | 93 | 1,138 | 93 | *< 0.001* |
| Paternal parental leave (yes) | 8,466 | 79 | 1,175 | 71 | 820 | 65 | *< 0.001* |
| Psychological risk assessment (yes) | 3,878 | 36 | 491 | 29 | 422 | 26 | *< 0.001* |
| *Antenatal education (number of sessions)* | *4* | *3* | *3* | *3* | *2* | *3* | *< 0.001* |
| **Sociodemographic** | | | | | | | |
| *Mother's age at birth* | *31* | *5* | *31* | *5* | *32* | *5* | *< 0.001* |
| *Father's age at birth* | 33 | 6 | 33 | 6 | 37 | 8 | < 0.001 |
| Father's employment (yes) | 10,421 | 94 | 1,596 | 91 | 1,471 | 84 | < 0.001 |
| Parity (≥ 1 other child) | 6,015 | 54 | 957 | 55 | 1,015 | 58 | < 0.001 |
| Marital status | | | | | | | < 0.001 |
|     Married | 4,665 | 42 | 952 | 54 | 1,313 | 75 | |
|     In civil union | 2,022 | 18 | 196 | 11 | 55 | 3 | |
|     Other | 4,395 | 340 | 607 | 35 | 382 | 22 | |
| **Pregnancy** | | | | | | | |
| Timing of pregnancy (satisfying) | 8,547 | 77 | 1,315 | 75 | 1,312 | 75 | < 0.001 |
| Previous PPD (yes) | 712 | 6 | 118 | 7 | 91 | 5 | 0.11 |
| Psychological difficulties during pregnancy (yes) | 1,304 | 12 | 234 | 13 | 240 | 14 | < 0.001 |
| *Number of prenatal visits* | 9 | 2 | 9 | 2 | 8 | 3 | 0.11 |
| **Child** | | | | | | | |
| Sex (boy) | 5,695 | 51 | 873 | 50 | 918 | 52 | 0.25 |
| Breastfeeding at birth | | | | | | | < 0.001 |
|     Exclusive | 6,876 | 62 | 1,171 | 67 | 1,290 | 74 | |
|     Non exclusive | 957 | 9 | 204 | 12 | 293 | 17 | |
|     No | 3,249 | 29 | 380 | 22 | 167 | 10 | |

Variables in italic correspond to quantitative variables.

**Table 2** –Proportions of the effects of an increase of one unit of socioeconomic position on the risk of postpartum depression, mediated by social support during pregnancy, ELFE Cohort Study 2011-2013

| Mediated Proportion of Social Inequalities of PPD by | Non-migrant<br>N = 11,082 (76%) | Second generation migrant<br>N = 1,755 (12%) | First generation migrant<br>N = 1,750 (12%) |
|---|---|---|---|
| Good partner relationship | 6% | NA | NA |
| Sufficient partner support | 8% | NA | 11% |
| Antenatal education | -9% | NA | NA |
| Total | 5% | NA | 11% |

*NA: not applicable if SEP (resp. SSP) is not associated with SSP (resp. PPD)

**Supplementary Figure 1** Confirmatory factor analysis for maternal socioeconomic position latent factor, ELFE Cohort Study 2011-2013, N = 16,513, crude analysis.

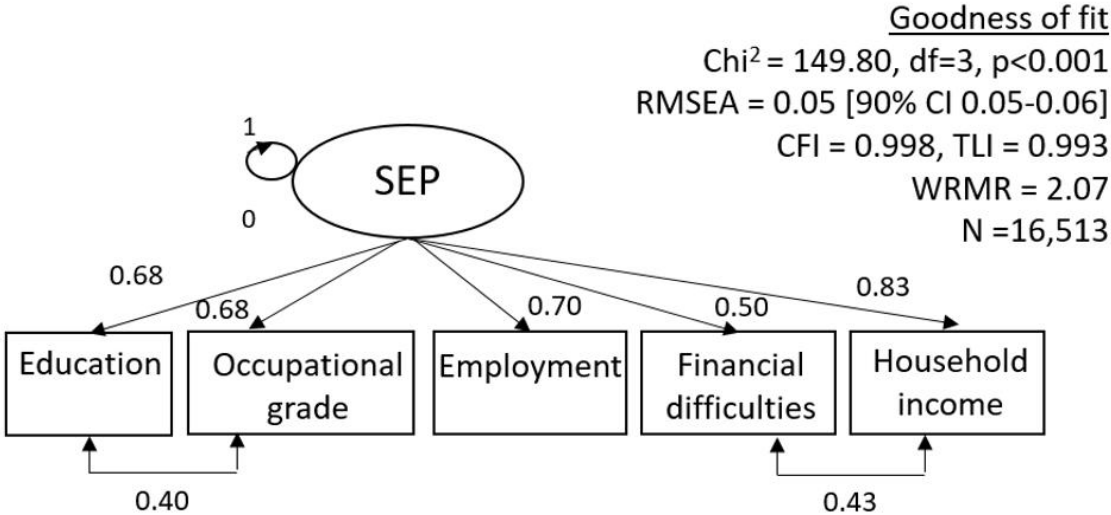

**Supplementary material**

**Supplementary Table 1 E-values for sensitivity analyses to unmeasured confounders**

| Independent -> Dependent variable | Non-migrant | Second generation migrant | First generation migrant |
|---|---|---|---|
| SEP -> Good partner relationship | 1.18 [1.00, +inf[ | NA | NA |
| SEP -> Satisfying partner support | 1.38 [1.31, +inf[ | 1.60 [1.43, +inf[ | 1.44 [1.22, +inf[ |
| SEP -> Antenatal education | 1.88 [1.83, +inf[ | 2.06 [1.90, +inf[ | 2.37 [2.21, +inf[ |
| SEP -> EPDS | 1.32 [1.25, +inf[ | 1.46 [1.25, +inf[ | 1.67 [1.29, +inf[ |
| Good partner relationship -> EPDS | 1.70 [1.56, +inf[ | 1.67 [1.21, +inf[ | NA |
| Satisfying partner support -> EPDS | 1.36 [1.21, +inf[ | NA | 1.50 [1.16, +inf[ |
| Antenatal education -> EPDS | 1.28 [1.21, +inf[ | NA | NA |

SEP = socioeconomic position, NA = not applicable (due to either not significant association between SEP and SSP or SSP and PPD). Analyses adjusted for mother's and father's age at child birth, parity, marital status, father's employment status, timing of pregnancy, psychological difficulties during pregnancy, previous postpartum depression, number of prenatal visits, child sex and breastfeeding at birth.